\documentclass[aps,prl,twocolumn,superscriptaddress,showpacs,floatfix]{revtex4}
\RequirePackage{amsmath}
\RequirePackage{amssymb} %\RequirePackage{bbm} 
\RequirePackage{xspace}
\newcommand{\dd}{\ensuremath{\text{d}}}
\usepackage{graphicx}

\newcommand{\kb}{\ensuremath{k_\text{B}}}

\newcommand{\etal}{\textit{et al.}\xspace}

\newcommand{\Tc}{\ensuremath{T_\text{c}}\xspace}

\newcommand{\Fig}[1]{Fig.~\ref{fig:#1}\xspace}
\newcommand{\SFig}[2]{Fig.~\ref{fig:#1}{#2}\xspace}
\newcommand{\SFigs}[4]{Figs.~\ref{fig:#1}{#2} and \ref{fig:#3}{#4}\xspace}

\newcommand{\Na}[1]{\ensuremath{^{#1}}\text{Na}}

\begin{document}
\title{Second sound dipole mode in a partially Bose-Einstein condensed gas}
\author{R.~Meppelink}
\affiliation{Atom Optics and Ultrafast Dynamics, Utrecht University,\\ P.O. Box 80,000, 3508 TA Utrecht, The Netherlands}
\author{S.~B.~Koller}
\affiliation{Atom Optics and Ultrafast Dynamics, Utrecht University,\\ P.O. Box 80,000, 3508 TA Utrecht, The Netherlands}
\author{J.~M.~Vogels}
\affiliation{Atom Optics and Ultrafast Dynamics, Utrecht University,\\ P.O. Box 80,000, 3508 TA Utrecht, The Netherlands}
\author{H.~T.~C.~Stoof}
\affiliation{Institute for Theoretical Physics, Utrecht University, \\P.O. Box 80,000, 3508 TA Utrecht, The Netherlands}
\author{P.~van der Straten}
\affiliation{Atom Optics and Ultrafast Dynamics, Utrecht University,\\ P.O. Box 80,000, 3508 TA Utrecht, The Netherlands}
\date{\today}

\begin{abstract} 
We study the second sound dipole mode in a partially Bose-Einstein condensed gas. This mode is excited by spatially separating and releasing the center-of-mass of the Bose-Einstein condensate (BEC) with respect to the thermal cloud, after which the equilibration is observed. The oscillation frequency and the damping rate of this mode is studied for different harmonic confinements and temperatures. The measured damping rates close to the collisionless regime are found to be in good agreement with Landau damping. For increasing hydrodynamicity of the cloud we observe an increase of the damping.
\end{abstract}
\maketitle

In 1938 Kapitza, and independently Allen and Misener, discovered that liquid $^4\text{He}$ below the $\lambda$-point can flow almost frictionless. Kapitza named this behavior superfluidity \cite{Nature.141.74,Nature.141.75}. Many of the properties of superfluid helium also appear in the gaseous Bose-Einstein condensates (BECs). In contrast to liquid helium, where the interatomic interaction is too strong to investigate the microscopic properties of superfluidity, the interactions in gaseous BECs are much weaker. The study of superfluid flow in dilute BECs can therefore deepen our understanding of superfluidity. 

The observation of vortices in a gaseous BEC \cite{PhysRevLett.84.806,PhysRevLett.83.2498} and the demonstration of the persistent flow of a BEC in a toroidal trap \cite{PhysRevLett.99.260401} give a striking demonstration of superfluidity. Evidence of the breakdown of superfluidity in a BEC is obtained in an experiment by Raman \etal, in which a blue detuned laser beam is moved through the BEC at different velocities \cite{PhysRevLett.83.2502}. Strong heating is observed only above a critical velocity $v_\text{c}$, which is found to be $v_\text{c}\simeq0.25c$, with $c$ the local speed of sound at the peak density of the BEC. 

In this Letter the flow of a BEC through a thermal cloud in a harmonic potential is studied by exciting a dipole oscillation of the BEC, whereas the thermal cloud initially remains stationary.  In the hydrodynamic regime this out-of-phase mode of the trapped Bose gas is the analog of the usual second sound mode in bulk superfluid helium \cite{PhysRevA.57.4695}. For this second sound dipole mode we study for the first time its frequency and damping rate from the collisionless to the hydrodynamic regime. In contrast to liquid helium, our analysis allows for a direct measurement of the position of the superfluid component (condensed atoms) with respect to the normal fluid (thermal atoms), which allows for an unequivocal determination of the second sound dipole mode. A dipole oscillation of the center-of-mass of the whole system is undamped in a harmonic potential and this in-phase dipole oscillation is often used to measure the trap frequency. 

The thermal cloud can also be tuned into the collisionless regime, in which the mean free path of the thermal atoms is larger than the axial size of the cloud. As we will show, the damping of collective excitations in a collisionless partly condensed BEC is primarily caused by Landau damping, i.e.~mean-field interactions mediate the transfer of energy from the condensate to the thermal cloud, leading to the damping of collective modes. Landau damping was first discussed by Landau in the context of the damping of plasma oscillation and plays a key role in a broad variety of fields, for instance the damping of phonons in metals, the damping of quarks and gluons in quark-gluon plasmas and the anomalous skin effect in metals.

Previously, some experiments have been performed in which the BEC and the thermal cloud move with respect to each other and either the BEC \cite{PhysRevLett.81.500} or the thermal cloud \cite{PhysRevA.66.011604} is initially at rest. In the experiment presented in Ref.~\cite{PhysRevA.66.011604} the thermal cloud is in the collisionless regime. In a pioneering experiment presented in Ref.~\cite{PhysRevLett.81.500} the thermal cloud is in the crossover region toward the hydrodynamic regime, but only one measurement series is presented. It therefore does not provide a study of the dependence of the superfluid flow on for instance the temperature or the collision rate. In contrast, the number of theoretical studies of this subject greatly outnumbers the available experimental work \cite{PhysRevA.57.4695,PhysRevA.62.053602,PhysRevA.65.011601,JLowTempPhys.116.277,PhysRevA.72.053630}.

The experimental setup, as described in detail in Ref.~\cite{RevSciInstr.78.013102}, is capable of creating Bose-Einstein condensates containing up to $3 \cdot 10^8$ sodium atoms, limited by three-body decay. In this setup we have reached the axially hydrodynamic regime in the thermal cloud above the transition temperature \Tc. The axially hydrodynamic regime is reached by evaporatively cooling atoms in an axially strongly decompressed trap with an aspect ratio of up to $1:65$ \cite{PhysRevA.75.031602}. The radial trap frequency is given by $\omega_\text{rad}/(2 \pi) = 95.56$~Hz. Here, we cool atoms to temperatures below $T_\text{c}$ for three different values of the axial confinement.

We introduce a measure for the hydrodynamicity of the thermal cloud in the axial direction $\tilde{\gamma}\equiv \gamma_{22}/\omega_\text{ax}$, where the collision rate $\gamma_{22} = n_\text{eff} \,\sigma\, v_\text{rel}$ is the average number of collisions in the thermal cloud. Here, $v_\text{rel}$ is the relative velocity, $\sigma =8\pi a^{2}$ is the isotropic cross-section of two bosons with $s$-wave scattering length $a$ and $n_\text{eff}=\int \dd \vec{r}\, n_\text{ex}^2(\vec{r})/\int \dd \vec{r} \, n_\text{ex}(\vec{r})$ is the effective density of the trapped thermal atoms with density $n_\text{ex}$. For our parameters, $\gamma_{22} \simeq 90 \text{s}^{-1}$ for the highest number of atoms and weakest axial confinement, which corresponds to a hydrodynamicity of  $\tilde{\gamma}\lesssim 10$. Note that the thermal cloud becomes less hydrodynamic for stronger axial confinement due to two effects: the increased three-body losses and the radial expansion. We have observed the crossover from the collisionless regime to the hydrodynamic regime by studying the heat conduction in a thermal cloud above \Tc \cite{PhysRevLett.103.095301}. Furthermore, it is noted in Ref.~\cite{PhysRevA.65.011601} that the thermal cloud is deeper in the hydrodynamic regime when a BEC forms due to collisions with condensed atoms. As a result the thermal cloud is even more hydrodynamic below \Tc than it is above the transition temperature.
\begin{figure}[htbp]
  \begin{center}
    \includegraphics[width=0.75\columnwidth]{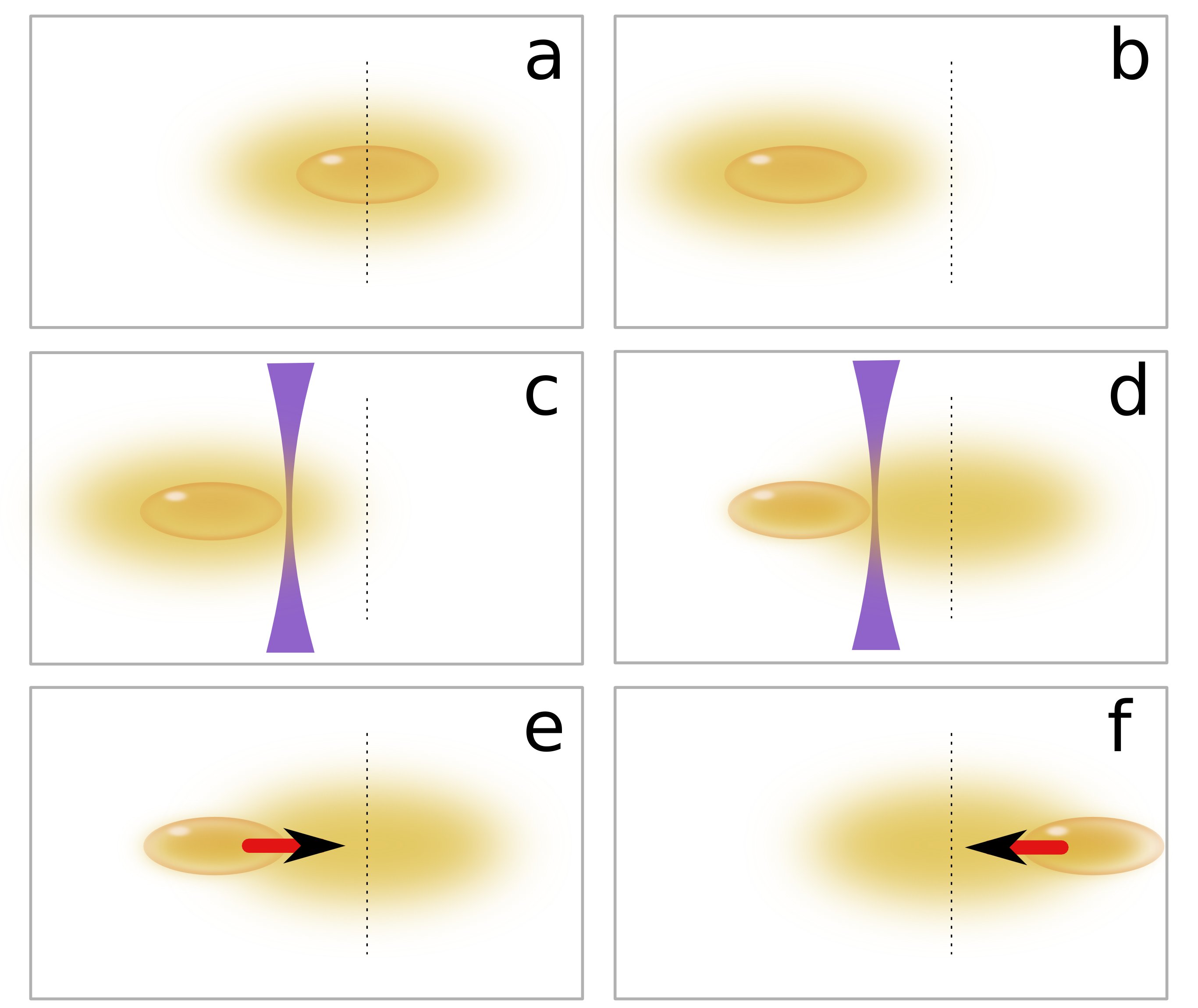}
  \end{center}
  \caption{Schematic representation of the excitation procedure of the second sound dipole oscillation. See text for more details.}% Fig. (a) shows a BEC in equilibrium at finite temperature, where the position $z=z_0$ of the center of the confinement is indicated by the dotted line. Next, the center of the trap is adiabatically displaced as is shown in Fig. (b) and the dipole beam gradually turned on as is shown in Fig. (c). The center of the trap is adiabatically displaced back to $z=z_0$, but only the thermal cloud can pass the dipole potential, as shown in Fig. (d). By suddenly removing the dipole potential the BEC will undergo a dipole oscillation, while the thermal cloud is initially at rest (Fig. (e) \& (f)).  }
  \label{fig:schemoop}
\end{figure}

The procedure used for the excitation of a second sound dipole mode, schematically shown in \Fig{schemoop}, is conducted as follows: initially, the center of the magnetic trap (MT) at $z=z_0$ (\SFig{schemoop}{a}) is displaced adiabatically in the axial direction over slightly more than the length of the BEC (\SFig{schemoop}{b}). Next, a focused blue-detuned laser beam is aligned perpendicular to the axial axis of the system at position $z_\text{d}$ with $z_0<z_\text{d}<z_1$, where it does not overlap with the BEC (\SFig{schemoop}{c}). This beam is detuned $25$ nm below the \Na{23} $\text{D}_2$-transition and its intensity is gradually ramped up to roughly $10^4$~mW/cm$^2$ in $100$ ms. This corresponds to a maximum potential of roughly $V_\text{dip} \simeq 1.1 \times \mu$, where $\mu$ is the chemical potential of the initial cloud. The intensity is continuously measured using a photo diode in order to stabilize the intensity using a feed-back circuit controlling the efficiency of an acousto-optical modulator, which deflects the light used for the dipole beam.  Next, the center of the trap is displaced adiabatically back to its original position $z=z_0$ (\SFig{schemoop}{d}). Since $\mu <V_\text{dip}<\kb T$, with $\kb$ the Boltzmann constant and $T$ the temperature, only the thermal cloud can pass the dipole potential and returns to $z=z_0$, while the condensate remains at $z=z_1$. Finally, the dipole beam is turned off and the BEC will undergo a dipole oscillation, while the thermal cloud is initially at rest (\SFigs{schemoop}{e}{schemoop}{f}). The distance $\Delta z = z_\text{d}-z_0$ determines the maximum velocity $ v = \omega_\text{ax} \Delta z$ of the center-of-mass of the condensate. The final position of the center of the MT is used to tune $ \Delta z$ and thus $v$.

\begin{figure}[htbp]
  \begin{center}
    \includegraphics[width=0.85\columnwidth]{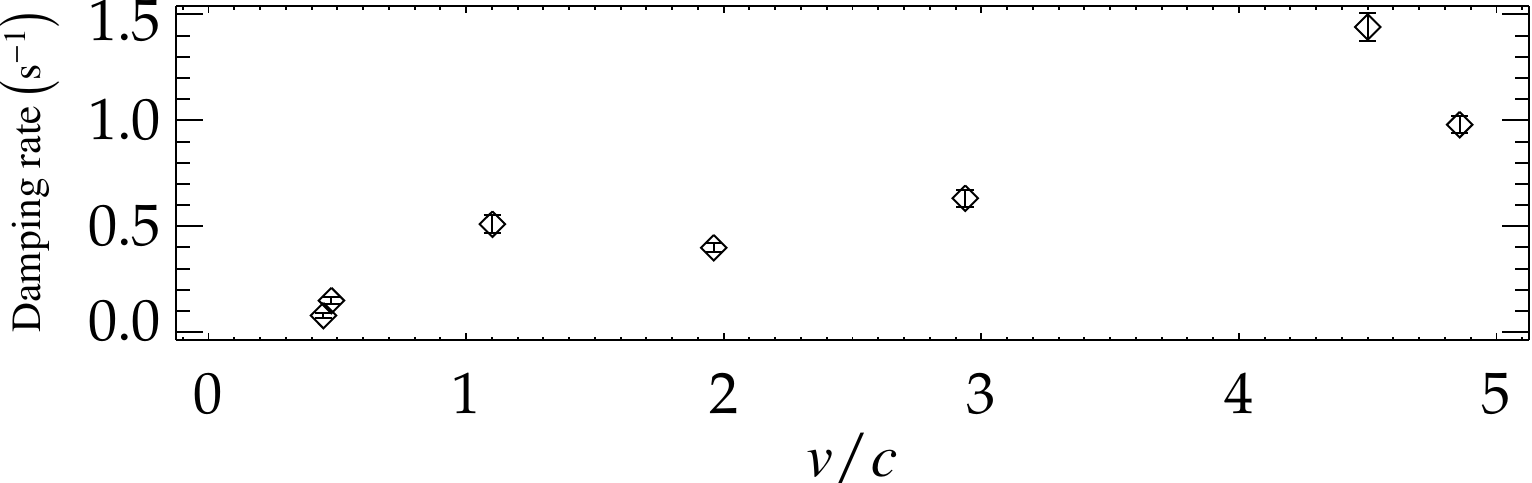}
  \end{center}
  \caption{The damping rate of the out-of-phase  oscillation as a function of the initial velocity $v/c$, where $c$ is the central speed of sound determined using the measured BEC density. The uncertainty in the measured damping rate and in $c$ is larger for high $v$ due to the depletion of the condensate.}
  \label{fig:critvel}
\end{figure}

We have excited the cloud for various $v$ in order to observe, if there is a threshold velocity $v_\text{c}$ above which the out-of-phase oscillation is strongly damped. The damping rate for different values of $v$ is shown in \Fig{critvel}. For $v > c$ we observe an increase of the damping rate of the out-of-phase oscillation for increasing $v$. The strong damping in this regime is accompanied by a large reduction of the condensate fraction, which eventually leads to the complete depletion of the BEC. For smaller $v$ (but still above $v_\text{c}$) we observe two distinct damping rates as long as the relative velocity is sufficiently reduced below $v_\text{c}$ before the condensate is fully depleted: first we observe a high damping rate, but as soon as the relative velocity is below $v_\text{c}$ a much lower damping rate is observed.  For excitations with $v \lesssim v_\text{c}$ we observe a low damping rate, independent of $v$ within the experimental accuracy. The threshold value depends on our trap geometry and is determined at $v_\text{c}/c\approx 0.59\pm0.05$ for the weakest axial confinement. For the strongest confinement used we find $v_\text{c}/c\approx 0.8\pm0.1$. The observation of a threshold velocity indicates the onset of a strong damping mechanism, which we consider as the breakdown of superfluidity. The experiments described in the remainder of this Letter are conducted in the regime of low damping ($v<v_\text{c}$).

A measurement series contains roughly $100$ shots at various hold time $\tau$ after the dipole beam is suddenly turned off. At the end of the hold time an absorption image is taken after the cloud is allowed to expand during a 85 ms time-of-flight. During a measurement series the situation before the dipole beam is turned off is monitored a few times to check for the absence of condensed atoms in the center of the trap. The thermal cloud is easily distinguished from the BEC in the absorption images due to the distinct density profiles. The final images are fitted to a bimodal distribution. This distribution is the sum of a Maxwell-Bose distribution modeling the thermal cloud and a Thomas-Fermi (TF) distribution modeling the BEC. The fit yields ten parameters; the axial and radial positions, the axial and radial sizes and the optical densities of both components. The axial position of both components is used to observe the dipole oscillations, the total absorption is used to determine the column density of both components, and the radial size of the thermal cloud is used to determine the temperature of the system. Furthermore, the aspect ratio of the condensate is used to detect the quadrupole oscillation induced by the excitation procedure, which turns out to be small. This mode does not couple to the dipole mode and is therefore not of interest to us here.  

\begin{figure}[htbp]
  \begin{center}
    \includegraphics[width=0.89\columnwidth]{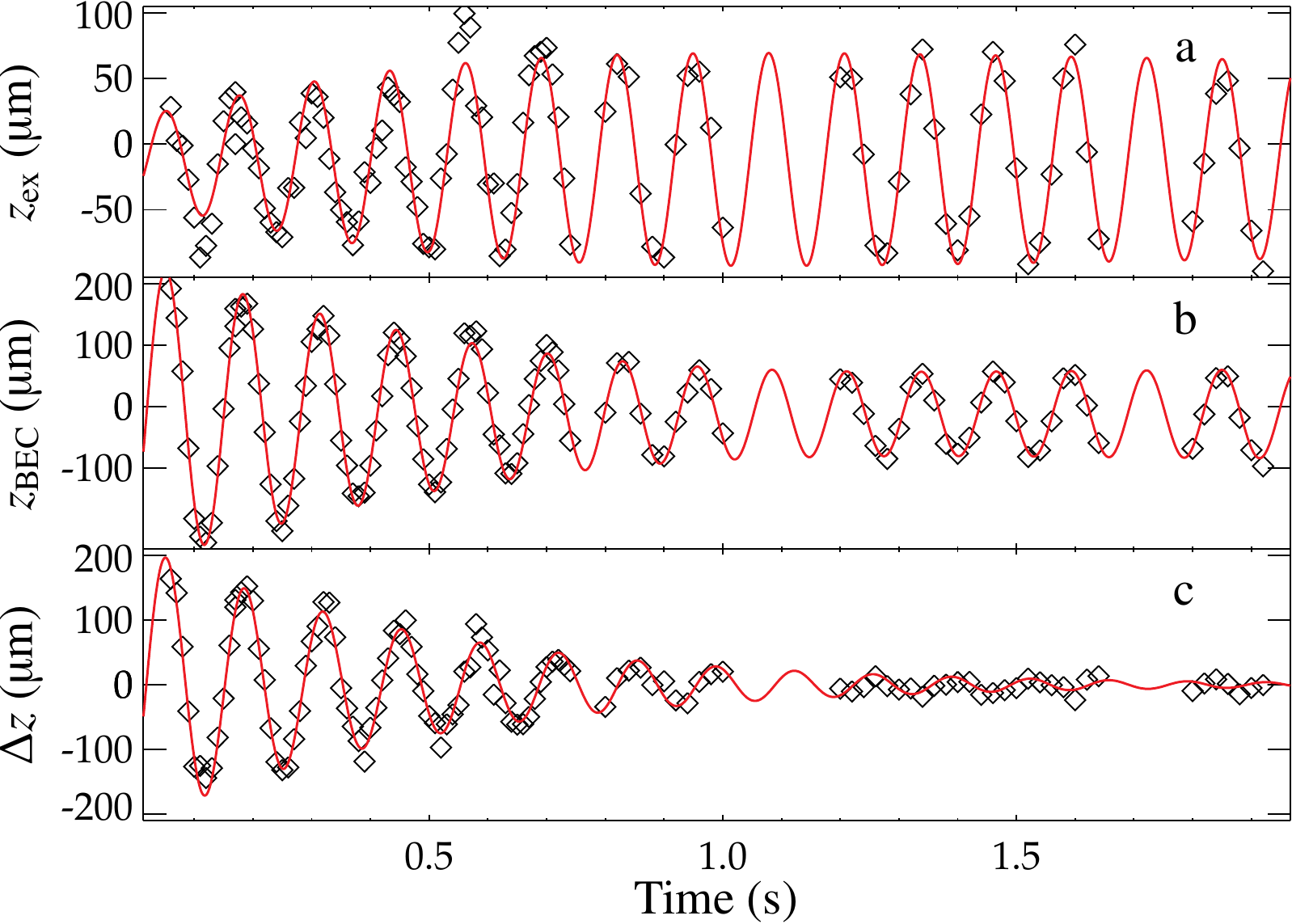}
    \caption{The in-phase and out-of-phase dipole oscillation of a system characterized by $T= 0.39 \mu$K and $\mu/h= 2.6$~kHz for $\omega_\text{ax}/(2\pi)= 7.78$~Hz. In Figs.~(a) and (b) the axial position $z_\text{ex}$ of the thermal cloud and $z_\text{BEC}$ of the BEC is plotted as a function of the hold time $\tau$. In Fig.~(c) the motion of the out-of-phase dipole oscillation is isolated by plotting $\Delta z$ as a function of $\tau$. The solid lines are the result of a fit to the data. Due to the destructive imaging scheme used each point represents the position of a newly prepared cloud.} 
\label{fig:oop8}
  \end{center}
\end{figure}
%\begin{figure}[htbp]
%  \begin{center}
%    \includegraphics[width=0.98\columnwidth]{graphics/oop_11.pdf}
%  \caption{The in-phase and out-of-phase dipole oscillation of a cloud, characterized by $T=\SI{0.46}{\microk}$ and $\mu/h=\SI{3.0}{kHz}$, for $\omega_\text{ax}/(2 \pi)=\SI{3.91}{Hz}$. For the description refer to the caption of \Fig{oop8}.}
%  \end{center}
%  \label{fig:oop4}
%\end{figure}
%\section{Results \& discussion}

A series is measured for three different values of the axial confinement and for two different temperatures. To compare the results under these different conditions, the temperature is adjusted in such a way that for a given temperature the initial chemical potential is roughly the same in all confinements. The axial position of the thermal cloud $z_\text{ex}$ and the BEC $z_\text{BEC}$  are plotted as a function of $\tau$ in \Fig{oop8}. Furthermore, the difference $\Delta z = z_\text{BEC}-z_\text{ex}$ is plotted to isolate the out-of-phase dipole oscillation from the in-phase oscillation. The thermal cloud, initially at rest, starts to oscillate when the BEC moves through it and eventually both the BEC and the thermal cloud oscillate simultaneously and in-phase, as can be seen in \Fig{oop8}. We refer to this final motion as the in-phase dipole oscillation. We have checked that the thermal cloud remains at rest as long as the dipole beam is present.

\begin{figure}[htbp]
  \begin{center}
    \includegraphics[width=0.75\columnwidth]{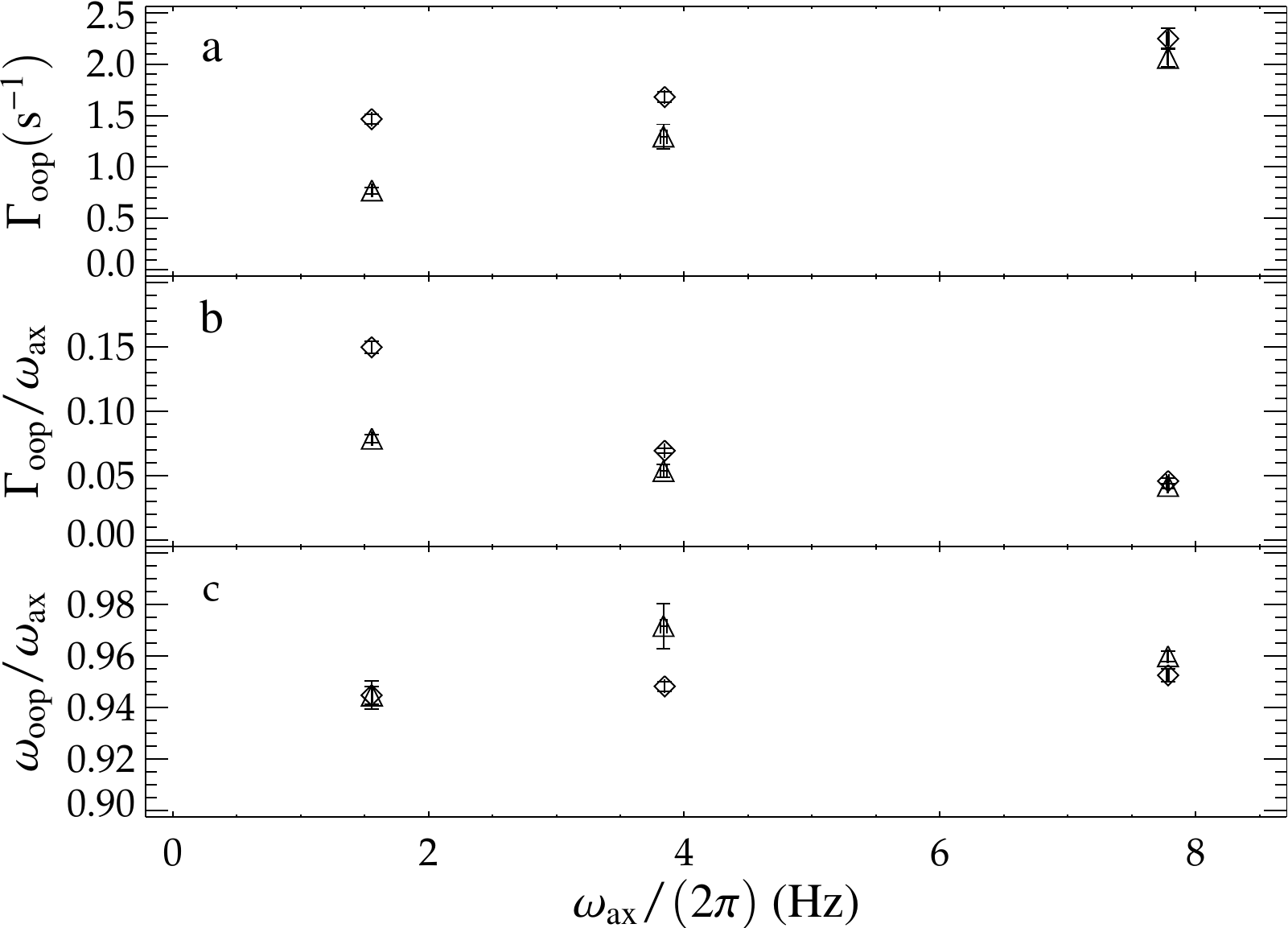}
  \end{center}
  \caption{The damping rate $\Gamma_\text{oop}$ (a), the normalized damping rate  $\Gamma_\text{oop}/\omega_\text{ax}$ (b), and the normalized oscillation frequency $\omega_\text{oop}/\omega_\text{ax}$ (c) as a function of the axial trap frequency $\omega_\text{ax}/(2\pi)$. The diamonds (triangles) indicates the high (low) temperature condition.} 
  \label{fig:oopvstrap}
\end{figure}

The positions $z_\text{ex}$ and $z_\text{BEC}$ are simultaneously fitted to a combination of two exponentially damped sinusoidal functions. The fit of the data yields the damping rate $\Gamma_\text{ip}$ ($\Gamma_\text{oop}$) and the frequency $\omega_\text{ip}$ ($\omega_\text{oop}$) of the in-phase (out-of-phase) dipole oscillation. The damping rate of the in-phase dipole oscillation does not deviate from zero within the experimental uncertainty, as is expected for a harmonic trap. The axial trap frequency determined from the in-phase oscillation frequency gives the same result as the trap frequency determination from a mutual center-of-mass oscillation ($\omega_\text{ip}=\omega_\text{ax}$). The quantities $\Gamma_\text{oop}$ and $\omega_\text{oop}$ are plotted as a function of the trap frequency in \Fig{oopvstrap}. %Furthermore, the results are tabulated in \Tab{tabel1}.

In \SFig{oopvstrap}{a} one can see that $\Gamma_\text{oop}$ decreases for decreasing axial trap frequency. However, $\Gamma_\text{oop}/\omega_\text{ax}$ increases for decreasing axial trap frequency as one can see in \SFig{oopvstrap}{b}. This indicates that an extra damping mechanism, which does not depend on the trap frequency, plays a significant role as well, especially for the highest $T$ and thus the most hydrodynamic clouds. 

%In addition to Landau damping collisional damping processes are expected in this case. In this regime the damping can be compared to models, which take collisional effects into account \cite{PhysRevA.62.053602,JLowTempPhys.116.277}. 

The normalized second sound oscillation frequency $\omega_\text{oop}/\omega_\text{ax}$ is shown in \SFig{oopvstrap}{c}. For all conditions $\omega_\text{oop}<\omega_\text{ax}$, where under the most hydrodynamic conditions $\omega_\text{oop}\approx 0.94 \omega_\text{ax}$. Furthermore, the largest deviations of $\omega_\text{oop}/\omega_\text{ax}$ from $1$ coincides with the largest $\Gamma_\text{oop}$ and is found for the most hydrodynamic clouds. The shift of $\omega_\text{oop}$ with respect to $\omega_\text{ax}$, up to 6\% in our experiment, is larger than the shift of 4.5\% observed in Ref.~\cite{PhysRevLett.81.500} and similar to the 6\% shift reported in Ref.~\cite{PhysRevA.66.011604}, although the reduced temperature $T/T_\text{c}$ is lower in our experiments. 

The mean-field potential of the thermal cloud causes the effective axial trapping potential to be slightly smaller. This effect reduces the effective trap frequency with roughly 2\% for the highest thermal density in these experiments and cannot account for the shifts of up to 6\%. As a consequence, the measured frequency shift and damping rate of the out-of-phase dipole mode both reflect that collisional effects play a role. Since $\omega_\text{oop}$ and $\omega_\text{ax}$ are determined simultaneously and with a high accuracy the frequency shift gives an accurate measure for these effects.

\begin{figure}[htbp]
  \begin{center}
    \includegraphics[width=0.75\columnwidth]{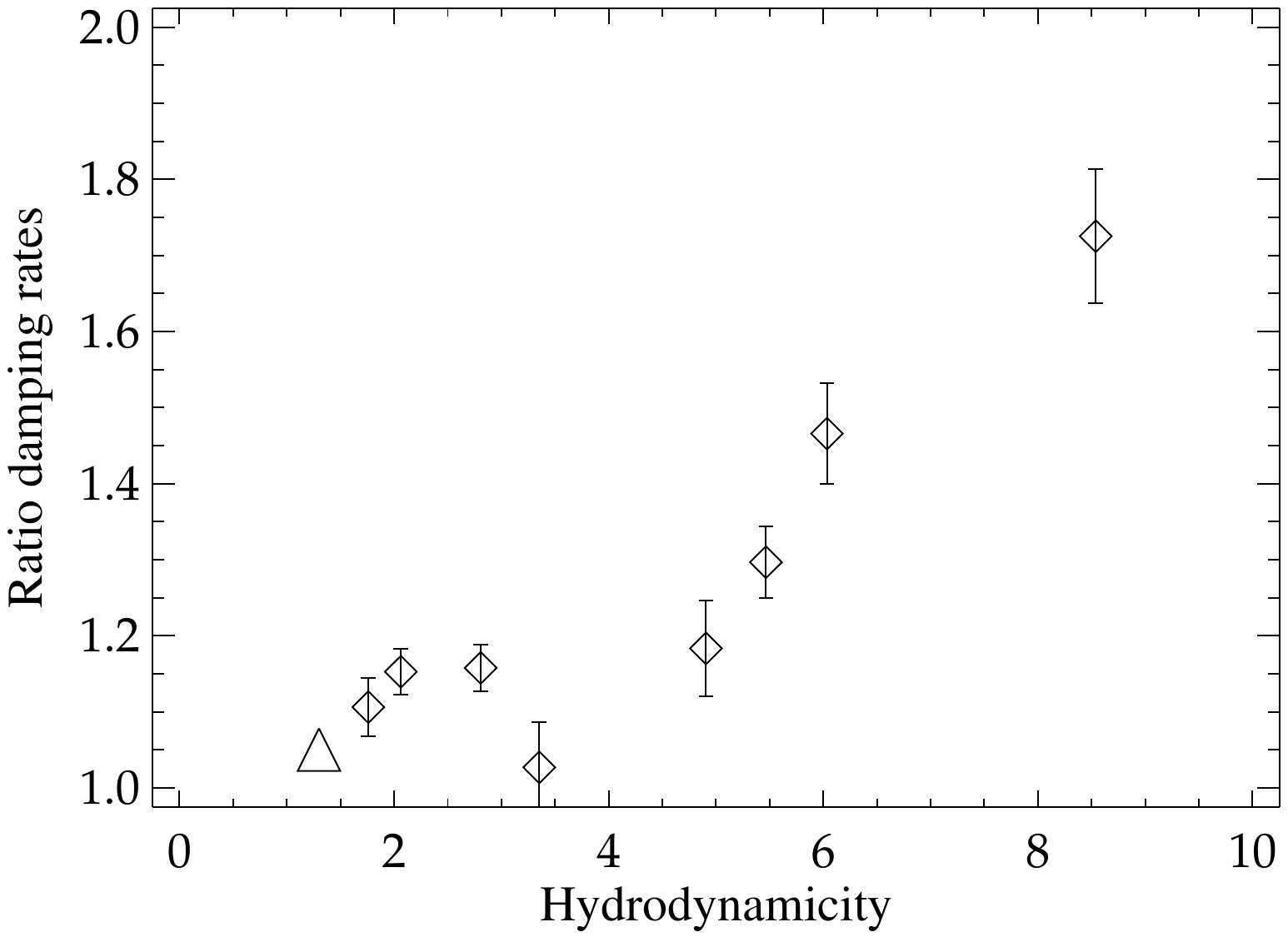}
  \end{center}
  \caption{The ratio of the measured damping rate and Landau damping $\Gamma_\text{oop}/\Gamma_\text{L}$ as a function of the hydrodynamicity parameter $\tilde{\gamma}$. The data point represented by the triangle is the result of a measurement taken from Ref.~\cite{PhysRevLett.81.500}.}
  \label{fig:superlandau}
\end{figure}

The measured damping rates are compared in \Fig{superlandau} to Landau damping. Landau damping in a homogeneous, weakly interacting Bose gas has been described in the collisionless limit by several authors \cite{AnnPhys.34.291,AnnPhys.82.1}. Landau damping increases with temperature, because of the larger number of  particle-hole pairs available at thermal equilibrium \cite{RevModPhys.71.463}. In our case, due to the confinement both the BEC and the thermal cloud have an inhomogeneous density distribution. Landau damping in such a system is determined by the condensate boundary region, and the result for the damping rate is different from that in a spatially homogeneous gas \cite{PhysRevLett.79.4056,PhysLettA.235.398,PhysRevLett.80.2269}. 
The damping rate for an inhomogeneous gas is given by $\Gamma_\text{L} =  \mathcal{C}\, \Gamma_\text{Lh} = \mathcal{C}\, 3 \pi \kb T \, a \, \omega/(8 \hbar c)$, where the speed of sound is given by $c=\sqrt{\mu/m}$, with $m$ the atomic mass \cite{PhysRevLett.80.2269}. Furthermore, $\omega$ is the frequency of the excitation, $\Gamma_\text{Lh}$ is the Landau damping rate for a homogeneous gas and $\mathcal{C}=12 c\,m /(\pi^{3/2}\,\kb\,\sqrt{m\,\hbar \omega}\,\text{log}\left[ 2 c^2 m/(\hbar \omega) \right])$ is a numerical coefficient to incorporate the inhomogeneity. Note that in our experiment $\Gamma_\text{L}$ is roughly independent of the speed of sound, since $\mu$ and therefore the central density is roughly constant in the different axial confinements. 

\Fig{superlandau} shows that we have measured the damping rate of the second sound dipole mode in a range from close to the collisionless regime up to the hydrodynamic regime with a hydrodynamicity close to 10. The ratio $\Gamma_\text{oop}/\Gamma_\text{L}$ increases strongly as a function of the hydrodynamicity $\tilde{\gamma}$ and this may serve as a stringent test for theoretical models, which attempt to describe the damping of hydrodynamic excitations in the Landau two-fluid (normal and superfluid) model. It shows that collisional processes, which have not been incorporated in the descriptions of Landau damping  discussed above, play an important role in damping the second sound dipole mode.

%For small $\tilde{\gamma}$ the measured damping rates are in good agreement with $\Gamma_\text{L}$. Therefore, we conclude that the damping of the out-of-phase dipole oscillation in the collisionless regime is caused by Landau damping. For higher values of $\tilde{\gamma}$ Landau damping underestimates the measured damping rates. In this regime, presumably also collisional damping plays a role. 

%\section{Conclusion \& Outlook}
In conclusion, we successfully excited a second sound dipole mode in a trapped Bose-condensed gas at finite temperature. The damping rate and the oscillation frequency of the second sound dipole mode is measured as a function of the axial confinement and the temperature with a high accuracy. We find that Landau damping, when the inhomogeneous density of the cloud is taken into account, shows good agreement to the measured damping rates close to the collisionless regime. In the hydrodynamic regime the measured damping rates are larger than the rates based on Landau damping. We hope that the measurements presented here encourages calculations based on models that take collionsional damping into account  and can provide a fruitful testing ground for such models.

This work is supported by the Stichting voor Fundamenteel Onderzoek der Materie ``FOM'' and by the Nederlandse Organisatie voor Wetenschaplijk Onderzoek ``NWO''.

%\bibliographystyle{h-physrev3}
%\bibliography{thesis}

\end{document}